\documentclass[12pt]{iopart}
\usepackage{graphicx}
\usepackage{iopams}

\newcommand{\lmk}{\left(}
\newcommand{\rmk}{\right)}
\newcommand{\lkk}{\left[}
\newcommand{\rkk}{\right]}

\begin{document}

\begin{flushright}
\hfill RESCEU-7/08, UTAP-596
\end{flushright}

\title{
Probing reheating temperature of the universe with gravitational wave background
}

\author{Kazunori Nakayama$^1$, Shun Saito$^2$, Yudai Suwa$^2$, 
and\\ Jun'ichi Yokoyama$^{3,4}$} 
\address{
$^1$Institute for Cosmic Ray Research, The University of Tokyo,
Kashiwa, Chiba 277-8582, Japan}
\address{
$^2$Department of Physics, Graduate School of Science, The 
University of Tokyo, Bunkyo-ku,
Tokyo 113-0033, Japan
}
\address{
$^3$Research Center for the Early Universe (RESCEU), 
Graduate School of Science, The University of Tokyo, Bunkyo-ku,
Tokyo 113-0033, Japan
}
\address{
$^4$Institute for the Physics and Mathematics of the Universe (IPMU), 
The University of Tokyo,
Kashiwa, Chiba 277-8568, Japan
}

\eads{\mailto{nakayama@icrr.u-tokyo.ac.jp}}

\date{\today}

\begin{abstract}
Thermal history of the universe after big-bang nucleosynthesis (BBN) 
is well understood both theoretically and observationally, and 
recent cosmological observations also begin to reveal the inflationary 
dynamics. 
However, the epoch between inflation and BBN is scarcely known. 
In this paper we show that the detection of the stochastic 
gravitational wave background around 1Hz provides useful information 
about thermal history well before BBN. 
In particular, the reheating temperature of the universe 
may be determined by future space-based laser interferometer experiments such as 
DECIGO and/or BBO if it is around $10^{6-9}$ GeV, 
depending on the tensor-to-scalar ratio $r$ 
and dilution factor $F$. 
\end{abstract}

\maketitle

\section{Introduction}
\label{sec:intro}

Recent cosmological observations have determined cosmological parameters 
with unprecedented accuracy \cite{Spergel:2006hy}. 
From measurements of cosmic microwave background (CMB) anisotropy, 
the baryon content of the universe $\Omega_b$ is determined 
and it agrees with the prediction of big-bang nucleosynthesis (BBN). 
This confirms the standard thermal history of the universe at least 
for $T\lesssim O(1)~$MeV, where BBN begins. 
On the other hand, the primordial power spectrum of the density 
perturbation has been revealed to be nearly scale invariant.
This has not only provided another evidence of inflation in the
early Universe \cite{lindebook}
but also given observational clues to study
its details such as the shape of the scalar
potential driving inflation \cite{yuragi}. 

Thermal history of the universe between inflationary era and 
BBN epoch, however, is least known observationally.
In general, the universe is dominated by the coherent oscillation 
of the inflaton soon after the inflation ends, 
and finally the inflaton decays into standard model particles, which 
realizes the hot radiation-dominated universe. 
Thus understanding this ``reheating'' stage of the universe is 
very important. In a simple standard inflation scenario, 
the reheating effect is characterized by one parameter, 
the reheating temperature $T_R$ \cite{Kolb:1990}, 
which is defined as 
\begin{equation}
  T_R = \left ( \frac{10}{\pi^2g_*(T_R)} 
      \right )^{1/4}\sqrt{\Gamma_{\phi}M_G},
  \label{TR}
\end{equation}
where $\Gamma_\phi$ is the decay rate of the inflaton,
$M_G=(8\pi G)^{-1/2}=2.4\times 10^{18}$~GeV is the reduced Planck scale 
and $g_*$ denotes the relativistic effective degrees of freedom. 
However, $T_R$ is hardly constrained from cosmological observations by now. 
Only the requirement is that $T_R$ must be larger than a few MeV \cite{Kawasaki:1999na} in order 
that light elements must be created in a usual way.\par

However, the reheating temperature $T_R$ contains rich information 
from the viewpoints of particle physics. First, as is obvious from 
Eq.~(\ref{TR}), $T_R$ is determined by the inflaton decay rate, 
and it depends on the inflaton properties, such as its mass, potential 
and interaction strength with standard model particles. 
Thus determining $T_R$ may have impacts on choosing realistic inflation 
models and inflaton candidates. 
Moreover, in supersymmetry (SUSY) \cite{Martin:1997ns}, which is 
well-motivated physics beyond the standard model, 
theoretical upper bounds on the reheating temperature is imposed 
from the so-called gravitino problem 
\cite{Khlopov:1984pf,Kawasaki:2004yh,Jedamzik:2006xz,Moroi:1993mb}. 
The abundance of the gravitino, which is the superpartner of the graviton, 
is bounded from above in order not to destroy light elements through 
its decay processes for an unstable gravitino, or not to overclose 
the universe for a stable gravitino. This gives upper bounds on $T_R$, 
since gravitinos are produced efficiently in the reheating era and 
its abundance is proportional to $T_R$.\par

Then, how can we probe such an early stage of the universe?
As is well known, observed CMB photons come from the last scattering 
surface located at redshift $z\sim 1100$. 
The universe is opaque to photons beyond the last scattering surface, 
and hence we cannot directly look over the universe of 
$z \gtrsim 1100$ by observations of photons. 
Even the cosmic background neutrinos cannot be used as a probe of 
the early universe with $T\gtrsim 1$~MeV, 
since neutrinos couple with nucleons strongly in such a high temperature 
environment. 

The universe, however,  is transparent to gravitational waves 
up to the Planck epoch in principle. This opens up the window for 
probing the very early universe, in particular the reheating epoch, 
with observations of gravitational waves.
Along with many species of unwanted relics, inflation dilutes
preexistent gravitational waves at the Planckian epoch.
Instead quantum mechanical
gravitational waves are produced in the inflationary era 
with an almost scale invariant power spectrum \cite{staro}. 
Although the amplitude of gravitational waves is constant in the
super horizon regime,
once a mode enters the horizon, it is reduced 
as the universe expands. Since the expansion rate depends on the 
equation of state of the universe, corresponding thermal history of 
the universe is imprinted in the gravitational wave spectrum at present.
Although much work has been done
 which treats the spectrum and detection 
possibility of stochastic gravitational wave background of inflationary 
origin in the literature 
\cite{Maggiore:1999vm,Allen:1987bk,Sahni:1990tx,Allen:1997ad,
Turner:1993vb,Turner:1993xz,Liddle:1993zj,Turner:1995ge,
Turner:1996ck,Seto:2003kc,Ungarelli:2005qb,Smith:2005mm,
Boyle:2005se,Smith:2006xf,Chongchitnan:2006pe,Friedman:2006zt,
Zhao:2006is,Watanabe:2006qe,Chiba:2007kz}, 
the previous work did not take into account the modification of 
the gravitational wave spectrum around the reheating epoch and assumed 
nearly flat spectrum above the observationally interesting 
frequency $\sim1$Hz, except for a few works where
it was claimed that the equation of state of the early universe can be probed 
by looking at the spectrum of the gravitational wave background \cite{Seto:2003kc}
(this kind of study was also done and extended in \cite{Boyle:2005se}).
Recently, we have pointed out that taking into account the reheating effect 
on the gravitational wave spectrum is essential both for the detection 
itself and the purpose of probing thermal history of the universe between 
BBN and inflation, with an emphasis on its impacts on particle physics
\cite{Nakayama:2008ip}.\par

In this paper, we give more detailed analyses of reheating effects 
on the gravitational wave background and show that future mission 
concepts based on space laser interferometer, like 
 Japan's DECIGO \cite{Seto:2001qf} and NASA's Big Bang 
Observer (BBO), may have a chance 
to detect primordial gravitational wave background produced in 
the inflatonary era and determine the thermal history of the universe 
before BBN epoch, in particular the reheating temperature $T_R$. 
We also discuss that some ``non-standard'' cosmological scenarios, 
including late-time entropy production, can be probed. \par

This paper is organized as follows.
In Sec.~\ref{sec:GW} the spectrum of primordial gravitational wave 
backgrounds, taking into account the modification 
around the reheating epoch, is derived.
Using this result, we discuss future prospects of the determination 
of the reheating temperature with space-based laser interferometers 
in Sec.~\ref{sec:reheating}. 
We discuss the case of late-time entropy production 
in Sec.~\ref{sec:entropy}. In Sec.~\ref{sec:implication}, 
the impacts of determining the reheating temperature on 
particle physics models are summarized. 
Sec.~\ref{sec:conclusion} is devoted to our conclusions.

\section{Primordial gravitational wave spectrum}
\label{sec:GW}

Primordial gravitational waves produced in the inflationary era 
have nearly scale-invariant spectrum and have effects on 
both large- and small-scale cosmological observations 
\cite{Maggiore:1999vm}. 
On large scales at observable frequeincies $\sim 10^{-16}$~Hz, 
the tensor metric perturbation generates B-mode polarization anisotropy 
of cosmic microwave background (CMB) for small multipole $l$, 
and such signal may be probed by next-generation ground-based or 
satellite 
observations of CMB (polarization) 
anisotropy \cite{Bock:2006yf,Amarie:2005in,Saito:2007kt}. 
On small scale, gravitational waves may be detected 
by future space-based interferometer experiments with frequency 
around $\sim 1$Hz. Thus primordial gravitational waves give 
much information on wide range of the cosmological scales. 
In particular, direct detection of gravitational waves may reveal 
the state of the universe well before BBN, as we will see. \par

Gravitational wave is described by the tensor perturbation 
on the metric, $h_{ij}$, which is defined as
\begin{equation}
	ds^2 = a^2(t)\left[
	-d\tau^2 + (\delta_{ij}+2h_{ij})dx^i dx^j \right],
\end{equation}
where $h_{ij}$ is symmetric under the exchange of $i$ and $j$,
and satisfies the traceless and transverse condition, 
${h_i}^i=0, {h_{ij}}^{,j}=0$. Thus it has two physical 
degrees of freedom, which we denote as $+$ and $\times$. 
Conformal time $\tau$ is defined as $d\tau =dt/a(t)$. 
Since tensor perturbation is gauge-invariant as it is, 
we do not need to care about gauge ambiguity as far as 
only the tensor mode is considered. 
The tensor perturbation $h_{ij}$ is expanded in the Fourier space as
\begin{equation}
	h_{ij}=\sqrt{8\pi G}\sum_{\lambda=+,\times} 
	\int \frac{d^3k}{(2\pi)^{3/2} }
	h_k^{\lambda} e_{ij}^{(\lambda)} e^{i{\bold {kx}}},
\end{equation}
where the superscript $\lambda$ denotes each polarization degree, 
$\lambda=+/\times$, and $e_{ij}^{(\lambda)}$ is the polarization tensor, 
which satisfies $e_{ij}^{(\lambda)}e^{ij(\lambda^\prime)}
=\delta^{\lambda \lambda^\prime}$. \par

Primordial gravitational waves are produced in the inflationary epoch 
with an almost scale-invariant spectrum. 
The amplitude of the produced gravitational wave is proportional 
to the Hubble scale during inflation, $H_{\rm inf}$. 
In terms of the dimensionless power-spectrum, 
it is given by \cite{Liddle&Lyth}
\begin{eqnarray}
	\Delta_h^{(\rm p)}(k)^2=& 64\pi G \left ( \frac{H_{\rm inf}}{2\pi} \right )^2
	\lkk 1-2\epsilon\ln\frac{k}{k_*}
	+2\epsilon(\eta-\epsilon)\lmk\ln\frac{k}{k_*}\rmk^2 \right ], 
\end{eqnarray}
counting two polarization states of the gravitational wave, 
where $\epsilon$ and $\eta$ denote 
the slow-roll parameters during inflation defined as 
$\epsilon = M_G^2(V^\prime/V)^2/2$ and $\eta = M_G^2(V^{\prime \prime}/V)$
with an inflaton potential $V[\phi]$ and its first and second derivative 
with respective to the inflaton field $\phi$, 
$V^\prime$ and $V^{\prime \prime}$. 
We take the pivot scale as $k_*=0.002$~Mpc$^{-1}$.
Here slow-roll parameters should be evaluated at the epoch when
CMB scale leaves the horizon. \par

On the other hand, the spectrum of the curvature perturbation is given by
\begin{eqnarray}
	\Delta_{\mathcal R}^2(k)=&\frac{4\pi G}{\epsilon}
	\left ( \frac{H_{\rm inf}}{2\pi} \right )^2
	\lkk 1+(-6\epsilon+2\eta)\ln\frac{k}{k_*} \right. \nonumber \\ 
	&\left. +
	(6\epsilon^2-4\epsilon \eta+2\eta^2-\xi)\lmk\ln\frac{k}{k_*}\rmk^2 \rkk ,	
\end{eqnarray}
where $\xi \equiv M_G^4 V^{'}V^{'''}/V^2$.
Its normalization is already well determined observationally, 
as $\Delta_{\mathcal R}^2 \sim 2.0\times 10^{-9}$ 
on CMB scale \cite{Spergel:2006hy}. 
Thus the slow-roll parameter $\epsilon$ is related to 
tensor-to-scalar ratio $r \equiv \Delta_h^2/\Delta_{\mathcal R}^2 (k_*)$ 
through the relation
\begin{equation}
	r=16\epsilon.  \label{ratio}
\end{equation}
This shows that $r$ is in general a small parameter and 
the tensor contribution is also expected to be small compared to 
the density perturbation. 
Notice that measuring $r$ fixes the inflation scale as 
\begin{equation}
	V_{\rm inf}=3M_G^2H_{\rm inf}^2 
	\simeq (3.2\times 10^{16}~{\rm GeV})^4 r.
\end{equation}

Then let us derive the observable power spectrum of inflationary 
gravitational wave background. 
The evolution of the gravitational wave is described 
by the following equation,
\footnote{
Here we neglect the anisotropic stress.
Including an anisotropic stress modifies the gravitational wave spectrum 
for the present frequency $\lesssim 10^{-9}$~Hz due to the neutrino 
free streaming effect \cite{Weinberg:2003ur,Watanabe:2006qe},
but our concern is the frequency around $\sim 1$~Hz. 
Hence the following arguments are not modified. 
}
\begin{equation}
	\ddot h^{\lambda}_k +3H \dot h^{\lambda}_k 
	+\frac{k^2}{a^2}h^{\lambda}_k =0.
	\label{eqofm_grav}
\end{equation}
The amplitude of the gravitational wave 
with comoving wave number $k$ remains constant when the mode 
lies outside the horizon. However, once it entered the horizon, 
its amplitude begins to damp. 
For the mode which enters the horizon in the matter-dominated regime, 
the solution is written as 
\begin{equation}
	h_k^{\lambda}(\tau)=h_k^{\lambda ({\rm p})}
	\left( \frac{3j_1(k\tau)}{k\tau}\right),
\end{equation}
where $j_{\ell}$ denotes the $\ell$-th spherical Bessel function. 
In general, the solution in a power law background $a(t)\propto t^p$ 
can be written in the form
\begin{equation}
	h_k(\tau) \propto 
	a(t)^{\frac{1-3p}{2p}}J_{\frac{3p-1}{2(1-p)}}( k\tau ),
\end{equation}
with Bessel function $J_n(x)$. Another damping factor comes from 
the fact that the relativistic degrees of freedom 
$g_*$ does not remain constant and expansion rate is modified
from simple power law $a(t) \propto T^{-1}$ in the early universe. 
This effect gives the damping factor \cite{Zhao:2006is,Watanabe:2006qe}
\begin{equation}
	\left ( \frac{g_*(T_{\rm in})}{g_{*0}} \right )
	\left ( \frac{g_{*s0}}{g_{*s}(T_{\rm in})} \right )^{4/3},
\end{equation}
on the power spectrum of the gravitational wave, where $T_{\rm in}$ 
denotes the temperature at which the corresponding mode 
enters the horizon, given by
\begin{equation}
	T_{\rm in}\simeq 5.8\times 10^6~{\rm GeV}
	\left ( \frac{g_{*s}(T_{\rm in})}{106.75} \right )^{-1/6}
	\left ( \frac{k}{10^{14}~{\rm Mpc^{-1}}} \right ).
\end{equation}
Moreover, recent cosmological observations have revealed that the 
expansion of the present universe is accelerating due to the dominance 
of unknown energy density, called dark energy 
(here we assume cosmological constant for simplicity). 
This also gives suppression factor $\sim (\Omega_m/\Omega_\Lambda)^2$ 
on the power spectrum \cite{Turner:1993vb}.
As a result, the present gravitational wave spectrum 
per log frequency interval is written in the form
\begin{equation}
	\Omega_{\rm gw}(f)= \frac{k^2}{12H_0^2}\Delta_h^2(k),
	\label{GWspec}
\end{equation}
where
\begin{eqnarray}
	\Delta_h^2(k)=&\Delta_h^{({\rm p})}(k)^{2}
	\left ( \frac{\Omega_m}{\Omega_\Lambda} \right )^2
	\left ( \frac{g_*(T_{\rm in})}{g_{*0}} \right ) 
	\left ( \frac{g_{*s0}}{g_{*s}(T_{\rm in})} \right )^{4/3} \nonumber \\
	&\times \left (\overline{ \frac{3j_1(k\tau_0)}{k\tau_0} } \right )^2
	T_1^2\left ( x_{\rm eq} \right )
	T_2^2\left ( x_R \right ),
\end{eqnarray}
with a bar denoting the average over many periods. 
Here, $T_1(x_{\rm eq})$ denotes the transfer function, 
which connects the gravitational wave spectrum of the mode 
which enters the horizon before and after matter-radiation 
equality, $t=t_{\rm eq}$. It is calculated as \cite{Turner:1993vb}
\begin{equation}
	T_1^2(x_{\rm eq})=
	\left [1+1.57x_{\rm eq} + 3.42x_{\rm eq}^2 \right ], \label{T1}
\end{equation}
where $x_{\rm eq}=k/k_{\rm eq}$ and 
$k_{\rm eq}\equiv a(t_{\rm eq})H(t_{\rm eq})
= 7.1\times 10^{-2} \Omega_m h^2$ Mpc$^{-1}$.\footnote{
	Coefficients in the right hand side of (\ref{T1}) differ 
	from those of Ref.~\cite{Turner:1993vb}
	reflecting the difference of the definition of $k_{\rm eq}$. 
}

\begin{figure}[t]
 \begin{center}
   \includegraphics[width=0.5\linewidth]{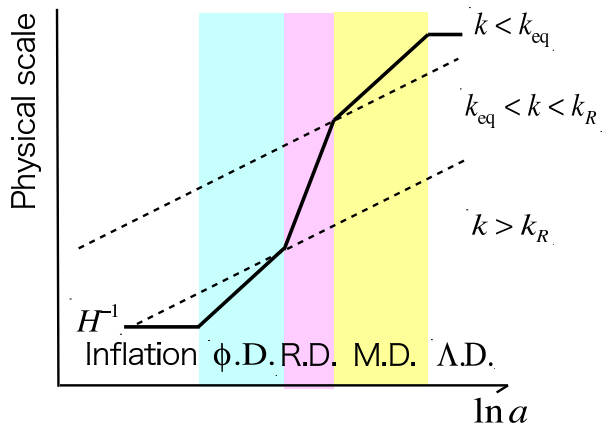} 
     \caption{A schematic picture of evolution of the Hubble 
	horizon scale $H^{-1}$ and physical wave length of some modes.
	$\phi$.D., R.D., M.D. and $\Lambda$.D. denote 
	the inflaton oscillation dominated era, radiation dominated era,
	matter dominated era, and cosmological constant dominated era, 
	respectively.}
      \label{fig:scenario}
 \end{center}
\end{figure}

On the other hand, $T_2(x_R)$ connects the mode which enters the horizon 
after and before the reheating ends. Because before the inflaton decays 
the universe is dominated by the coherent oscillation of the inflaton, 
the spectrum of the gravitational wave changes for the mode which 
enters the horizon at the inflaton-dominated epoch ($k > k_R$), where
\begin{equation}
	k_R\simeq 1.7\times 10^{13}~{\rm Mpc^{-1}}
	\left ( \frac{g_{*s}(T_R)}{106.75} \right )^{1/6}
	\left ( \frac{T_R}{10^6~{\rm GeV}} \right ).  \label{k_R}
\end{equation}
In terms of the frequency, this corresponds to
\begin{equation}
	f_R\simeq 0.026~{\rm Hz}
	\left ( \frac{g_{*s}(T_R)}{106.75} \right )^{1/6}
	\left ( \frac{T_R}{10^6~{\rm GeV}} \right ),  \label{f_R}
\end{equation}
which is close to the most sensitive frequency band of  DECIGO and BBO
for $T_R \sim 10^6~{\rm GeV}$. 
The transfer function $T_2(x_R)$ is obtained by solving simultaneously 
Eq.~(\ref{eqofm_grav}) and the following Friedmann equations 
taking the decay of inflaton into acount, 
\begin{eqnarray}	
	\dot \rho_\phi + 3H\rho_\phi = -\Gamma_\phi \rho_\phi,\\
	\dot \rho_r + 4H\rho_r = \Gamma_\phi \rho_\phi,\\
	H^2 = \frac{8\pi G}{3}(\rho_\phi +\rho_r),
\end{eqnarray}
where $\rho_\phi$ denotes the energy density of the inflaton 
coherent oscillation. 
We find that $T_2(x_R)$ is well approximated by 
\begin{equation}
	T_2^2(x_R)=\left [1-0.32x_R + 0.99 x_{R}^2 \right ]^{-1}, 
\end{equation}
where $x_R =k/k_R$. Thus $\Omega_{\rm gw}(f)$ behaves as 
$\propto f^{-2} (f^0)$ for the mode which enters in the horizon 
in the matter (radiation) dominated regime. 
In Fig.~\ref{fig:scenario} we show a schematic picture 
for the evolution of the Hubble horizon scale 
$H^{-1}$ by solid line, and physical wave length of various modes, 
$k>k_R$, $k_{\rm eq}<k<k_R$ and $k<k_{\rm eq}$ by dotted lines. 
Since the evolution of inflationary gravitational wave is 
sensitive to the expansion of the universe, $H^{-1}$, 
Fig.~\ref{fig:scenario} shows that characteristic feature 
of power spectrum can be probed for the mode around $f\sim f_{R}$. 

\begin{figure}[bp]
 \begin{center}
   \includegraphics[width=0.6\linewidth]{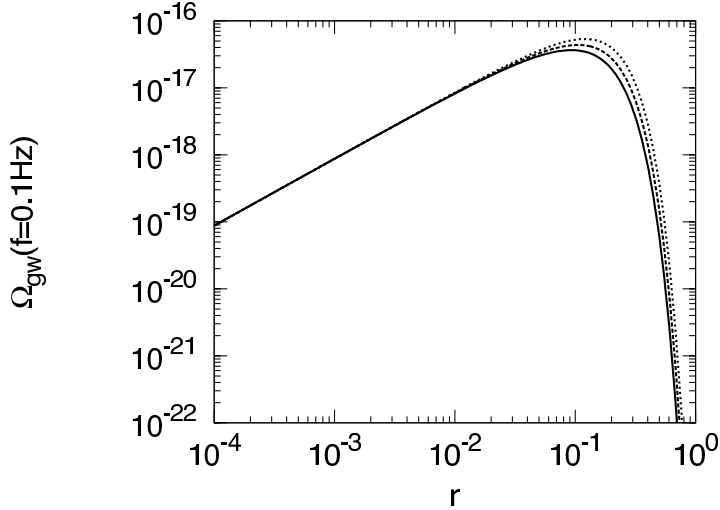} 
     \caption{$\Omega_{\rm gw}(f)$ at $f=0.1$Hz 
	for $\eta =0.01,0,-0.01$ from upper to lower.}
	\label{fig:f1Hz}
 \end{center}
\end{figure}

Before discussing the detection possibility, we briefly mention 
the effect of running spectral index. 
The running spectral index depends on the slow-roll parameters 
$\epsilon$ and $\eta$. 
Although $\epsilon$ is related to $r$ through 
the relation (\ref{ratio}), we are free to choose $\eta$ 
as long as the scalar spectral index $n_s (\simeq 1-6\epsilon +2\eta)$ 
lies in the favored range from recent WMAP results. 
In Fig.~\ref{fig:f1Hz} the resulting spectrum 
$\Omega_{\rm gw}(f)$ at $f=0.1$~Hz is shown for $\eta = 0.01,0,-0.01$ 
from upper to lower. 
Here sufficiently large reheating temperature $T_R \gtrsim 10^9$~GeV 
is assumed. As is clearly seen, for $r \gtrsim 0.1$ the gravitational 
wave amplitude at $f=0.1$~Hz decreases since the spectral tilt 
and its running becomes large. 
The difference of $\eta$ yields slight change 
on the gravitational wave amplitude for $r \lesssim 0.1$ 
(c.f. the current contraint on $r$ is $r<0.20$ (95\% C.L.)). 
Hereafter we set $\eta=0$ for simplicity.

\begin{figure}[tb]
 \begin{center}
   \includegraphics[width=0.6\linewidth]{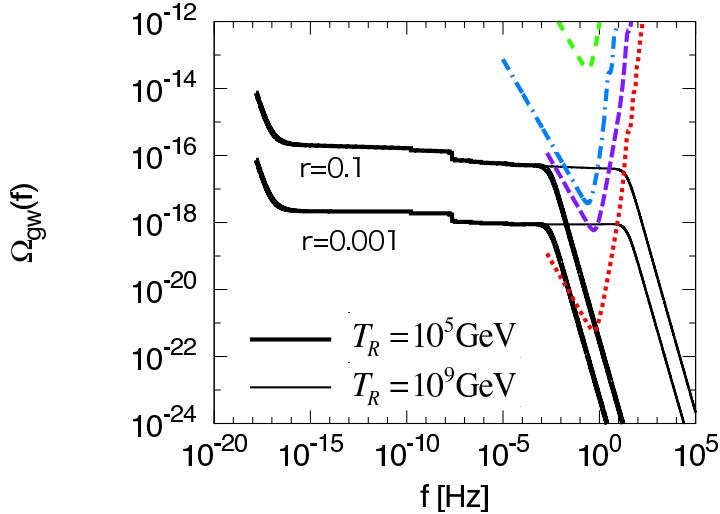} 
     \caption{Primordial gravitational wave spectrum for $T_R = 10^9$~GeV and 
     $T_R=10^5$~GeV are shown by thin and thick lines for $r=0.1$ and $0.001$. 
     Also shown are expected sensitivity of DECIGO (green dashed), 
     correlated analysis of DECIGO (blue dot-dashed), ultimate-DECIGO (purple dashed)
     and correlated analysis of ultimate-DECIGO (red dotted), from upper to lower.}
     \label{fig:GW}
 \end{center}
\end{figure}

In Fig.~\ref{fig:GW}, we show the resulting 
gravitational wave spectrum for $T_R = 10^9$ and $10^5$~GeV
with $r=0.1$ and $0.001$.
Together with the predicted gravitational wave spectrum, 
sensitivities of the future space-based laser interferometer experiments,
DECIGO, correlated analysis of DECIGO, ultimate-DECIGO (single)
and correlated analysis of ultimate-DECIGO \cite{Kudoh:2005as} 
are also presented. 
It can be seen that theoretical predictions are well above 
the sensitivity line of DECIGO (correlated) and ultimate-DECIGO 
for large enough $T_R$ and $r$, while the direct detection is 
undesirable for $T_R \lesssim 10^4$~GeV.

\section{Prospects for the determination of $T_R$ with 
future space-based laser interferometer experiments}
\label{sec:reheating}

The spectrum of the primordial gravitational wave background generated 
during inflation crucially depends on the reheating temperature $T_R$ 
after inflation, as can be seen from Fig.~\ref{fig:GW}.
Conversely, this fact opens up the possibility that future experiments 
devoted to detect gravitational wave background will probe the reheating 
stage of the universe.\par

The important parameters that determine the primordial gravitational 
wave spectrum are the tensor-to-scalar ratio $r$ and 
the reheating temperature $T_R$. Tensor-to-scalar ratio determines 
the overall normalization of the spectrum, and $T_R$ fixes the frequency 
above which the spectrum is significantly suppressed. 
The important point is that if the bending point of the gravitational 
wave spectrum determined by $T_R$ around the frequency 
given by (\ref{k_R}) lies above the sensitivity of detectors, 
$T_R$ can be determined by observations of gravitational waves. \par

The non-zero value of tensor-to-scalar ratio $r$ will be probed with 
measurements of the B-mode CMB polarization \cite{Amarie:2005in}. 
Since the B-mode is generated by the tensor perturbation only 
for large angular scale, $\ell\lesssim 100$, 
detection of the B-mode polarization indirectly confirms the 
existence of the gravitational wave background.
Notice that this effect is seen on very large-scale anisotropy 
with wavelength comparable to the present horizon scale, 
so the detection of the B-mode polarization is somewhat complementary 
to the direct detection of the gravitational wave on small scale with 
wavelength of the order of the Earth radius. 
The Planck satellite \cite{Planck}, scheduled to be launched in October 
2008, will measure $r$ up to $\sim 0.1$. 
On-going ground-based or balloon experiments devoted to 
detect CMB polarization, such as Q/U Imaging ExperimenT 
(QUIET) \cite{QUIET} and Clover \cite{Taylor:2004hha} 
will detect $r$ up to $\sim 0.01$. 
Ultimately future space mission will dedicate to detect the primary 
B-mode signal up to $r \gtrsim 10^{-3}$ \cite{Amarie:2005in}.
According to recent works \cite{Pagano:2007st,Smith:2008pf}, 
the observed value of the scalar spectral index, 
$n_s \simeq 0.961\pm 0.017$, implies somewhat large tensor contribution, 
$r\gtrsim 10^{-3}$ for various inflation models. Thus it seems to be 
plausible that non-zero $r$ will be confirmed by 
CMB polarization experiments. \par

Once $r$ is measured from those experiments, we may have a chance 
to detect signals of gravitational wave background of inflationary origin. 
Planned Laser Interferometer Space Antenna (LISA) \cite{LISA} cannot reach 
the required sensitivity even the largest possible value of $r$ 
allowed by WMAP5 and high enough $T_R$ are assumed. 
However, future mission concepts like  DECIGO and BBO are likely
to detect them or put a meaningful constraint on $T_R$. 
Therefore our main interest is to demonstrate the observable range of 
reheating temperature through such future experiments for various values 
of $r$. Before discussing results, however, it should be noticed that 
stochastic noise coming from some astrophysical processes must 
be taken into account for the purpose of detecting 
primordial gravitational wave background. 
In particular, gravitational waves from white-dwarf binaries 
are considered to  completely hide the primordial ones 
for the frequency $f \lesssim 0.1$Hz \cite{Farmer:2003pa}. 
But for the frequency range 0.1-10Hz, where the  DECIGO and BBO 
are most sensitive, foregrounds from astrophysical objects are separable.
\footnote{
Gravitational waves from the collapses of Population III stars 
may be a dominant contribution around the deci-Heltz band 
\cite{Buonanno:2004tp}. However, it crucially depends on 
the early star formation rate, which is very uncertain now, 
and such a signal would be separable if we adopt reliable 
abundance of Population III stars and take the duty cycle 
and their angular distribution into account.
}\par

\begin{figure}[tbp]
 \begin{center}
   \includegraphics[width=0.6\linewidth]{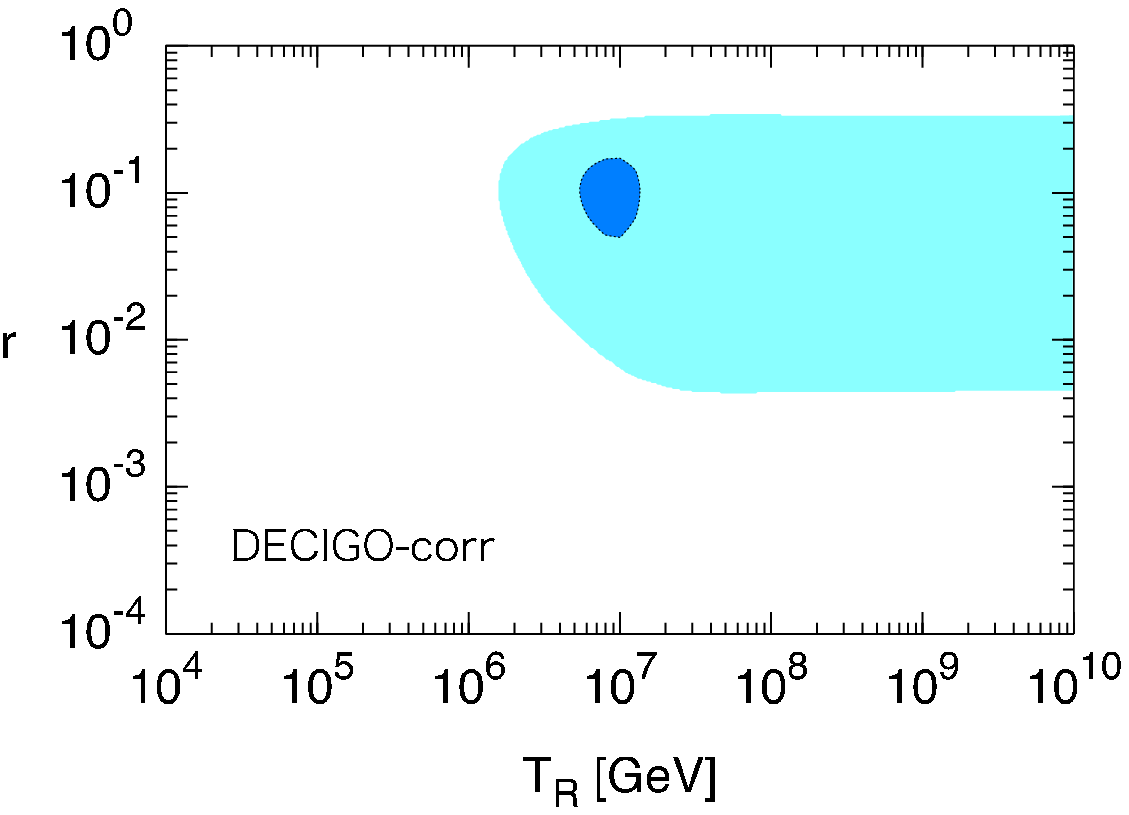} 
   \includegraphics[width=0.6\linewidth]{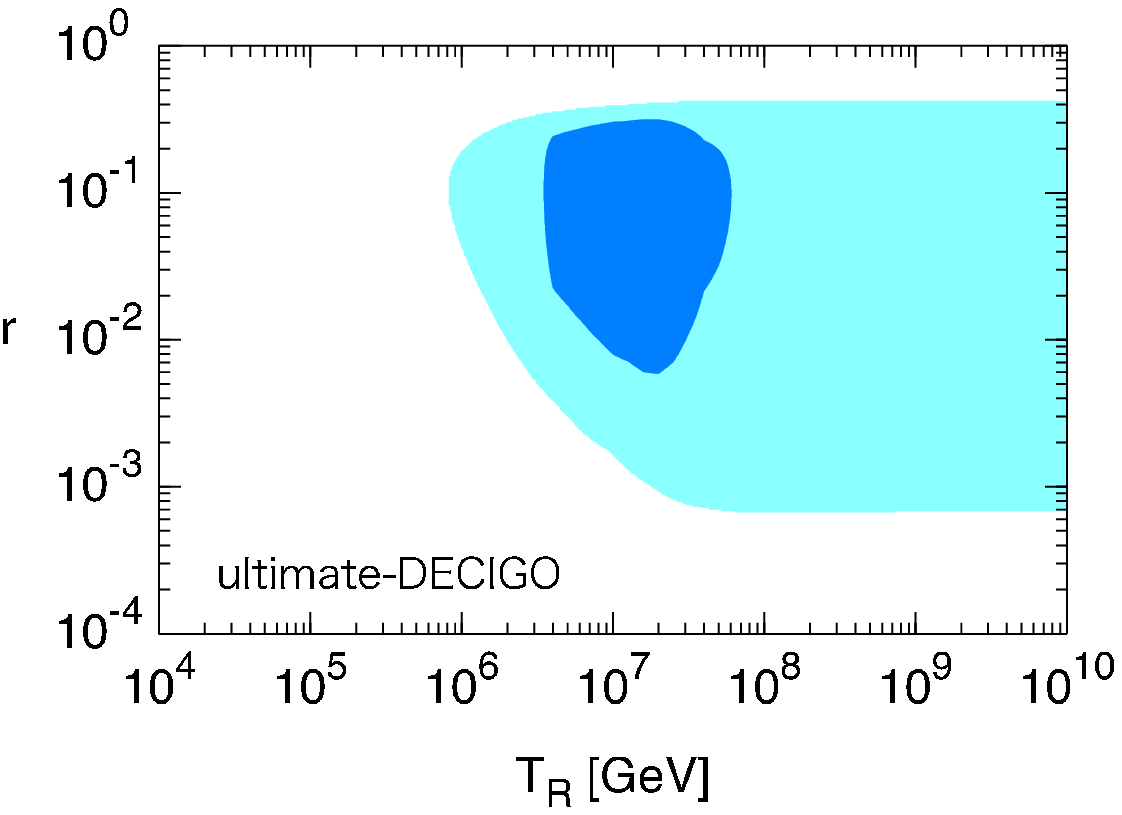} 
   \includegraphics[width=0.6\linewidth]{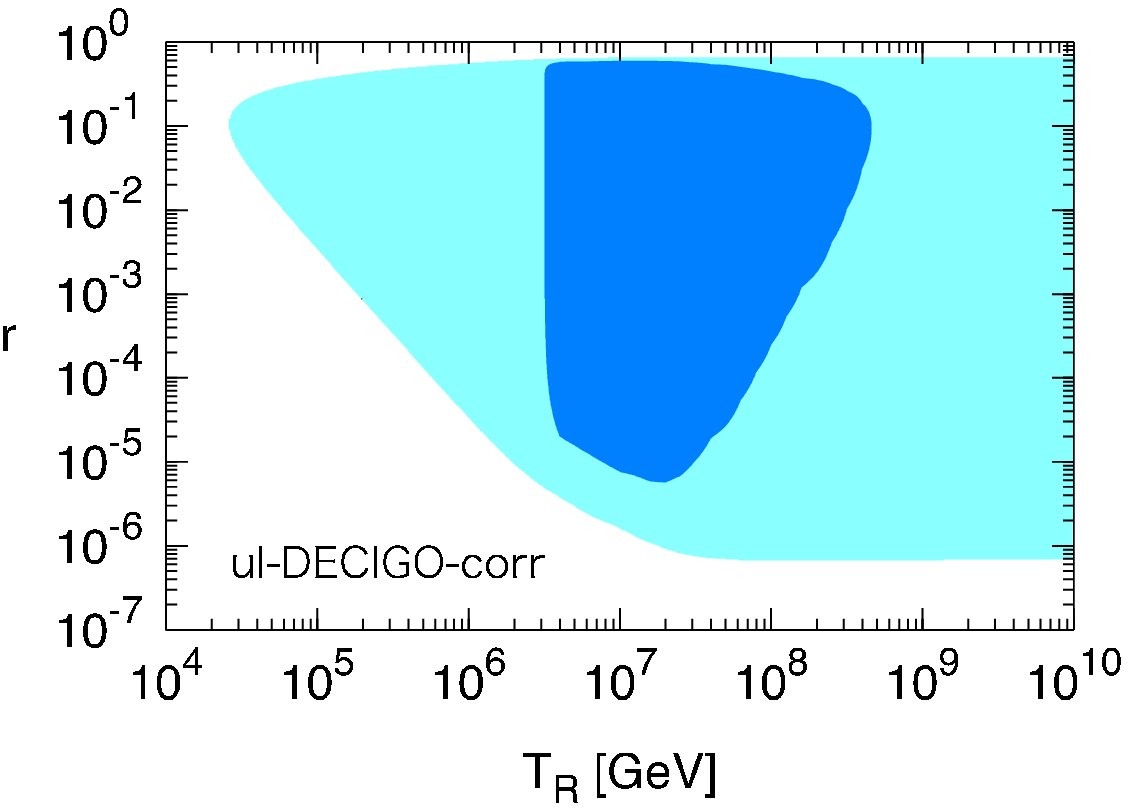} 
     \caption{
       In the outer light shaded region the gravitational wave background can be detected,
       and the inner blue shaded region shows the region where $T_R$ can be determined 
       with signal-to-noise ratio 5 by correlated analysis of DECIGO, ultimate-DECIGO (single)
       and ultimate-DECIGO (correlation) from upper to lower.
      }
     \label{fig:sensBBO}
 \end{center}
\end{figure}

In Fig.~\ref{fig:sensBBO}, we show parameter 
region on $r$-$T_R$ plane which will be probed with the 
correlated analysis of DECIGO, ultimate-DECIGO (single) and 
ultimate-DECIGO (correlation) respectively.
We have cut the sensitivity of these projects below $f\lesssim 0.1$~Hz,
taking into account the stochastic noise from white-dwarf binaries.
The gravitational wave background can be detected in the light blue shaded region. 
Furthermore the dark blue shaded region shows the parameter region where
the value of $T_R$ can be determined with signal-to-noise ratio 5. 
It is seen that for $10^{-3}\lesssim r \lesssim 1$, 
direct detection of gravitational wave background can determine 
the reheating temperature $T_R$, 
if it lies in the range $T_R\sim 10^{6}$-$10^8$~GeV. 

If higher $T_R$ is realized in nature, the 
inflationary gravitational wave background can be detected 
by DECIGO or BBO, but the value of $T_R$ remains undetermined. 
In this case, the lower bound on $T_R$ can be read off from these figures. 
For example if $r=0.1$, the detection of gravitational wave background 
means $T_R\gtrsim 10^8$~GeV, which provides useful constraints 
on some particle physics models, 
as we will see in Sec.~\ref{sec:implication}.

\section{Late-time entropy production}
\label{sec:entropy}

So far, we have assumed that there were no late-time 
entropy production processes after the completion of 
reheating after inflation. 
However, this may be too simplified assumption. 
Let us consider the case some scalar field $\chi$ other than the inflaton 
dominates the universe after the inflaton decays and $\chi$ 
eventually decays releasing huge entropy. 
Examples are Polonyi \cite{Moroi:1994rs} or moduli field 
\cite{Moroi:1999zb,Endo:2006zj}, 
scalar partner of the axion \cite{Kim:1992eu}, 
or others. 
Such late-decaying particles are interesting since they dilute 
cosmologically harmful gravitinos. 
We denote the dilution factor as $F$, defined by
\begin{equation}
	F =\frac{s(T_\chi)a^3(T_\chi)}{s(T_R)a^3(T_R)} 
	= \frac{T_R}{T_\chi}
	\left ( \frac{\rho_\chi}{\rho_\phi} \right )_{T_R},
\end{equation}
where $T_\chi$ is the decay temperature of $\chi$, 
which must be larger than a few MeV, 
and $\rho_\chi$ denotes the energy density of 
the $\chi$-field coherent oscillation.
\footnote{
Here we assume that $\chi$ begins to oscillate 
during the inflaton oscillation dominated phase.
}
For a simple model where the $\chi$-field has an initial amplitude 
$\chi_i$ and begins to oscillate when the Hubble parameter becomes 
equal to the mass of the $\chi$, it is estimated as
\begin{equation}
	F = \frac{T_R}{T_\chi}\frac{\chi_i^2}{3M_G^2}.
\end{equation}
The abundance of all dangerous cosmological relics produced 
in the reheating era after inflation, such as the gravitino 
discussed below, are diluted by this factor.\par 

\begin{figure}[t]
 \begin{center}
   \includegraphics[width=0.5\linewidth]{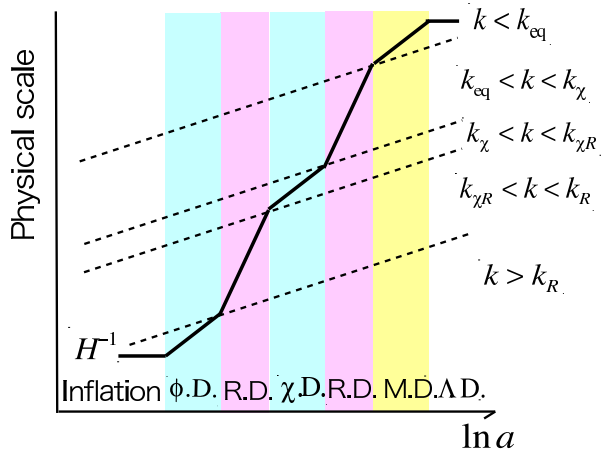} 
     \caption{A schematic picture of evolution of the Hubble horizon scale $H^{-1}$ and 
     physical wave length in the presence of late-time entropy production from $\chi$.
     $\chi.$D. represents $\chi$-dominated era.}
      \label{fig:scenario_F}
 \end{center}
\end{figure}

Importantly, such non-standard cosmological evolution scenarios 
are imprinted in the present gravitational wave spectrum. 
In the presence of such a late-decaying particle, 
the additional $\chi$-matter dominated era suppresses 
the gravitational wave amplitude for the frequency which reentered 
the horizon during or before $\chi$ begins to dominate the universe. 
(Fig.~\ref{fig:scenario_F} shows a schematic picture.) 
The spectrum now becomes 
\begin{equation}
	\Omega_{\rm gw}(f,F) = \Omega_{\rm gw}(f)
	\times T_2^2(x_\chi)T_1^2(x_{\chi R}),
\end{equation}
where $\Omega_{\rm gw}(f)$ is given by Eq.~(\ref{GWspec}) 
with $k_R$ replaced by
\begin{equation}
	k_R(F)=k_R F^{-1/3}.
\end{equation}
Here $x_\chi (\equiv k/k_\chi)$ corresponds to the wavenumber 
which enters the horizon at the decay of $\chi$, 
\begin{equation}
	k_\chi \simeq 1.7\times 10^{7}~{\rm Mpc^{-1}}
	\left ( \frac{g_{*s}(T_\chi)}{106.75} \right )^{1/6}
	\left ( \frac{T_\chi}{1~{\rm GeV}} \right ),  \label{k_chi}
\end{equation}
and $x_{\chi R} (\equiv k/k_{\chi R})$ corresponds to the epoch 
when $\chi$-domination begins, given by
\begin{equation}
	k_{\chi R}(F) = k_{\chi}F^{2/3}.
\end{equation}
We can see that for the mode $k_{\chi R}<k<k_R$, which corresponds 
to the mode which reenters the horizon in the radiation dominated era 
before the $\chi$-domination, the energy density of the gravitational waves 
is suppressed by the factor 
$\sim (k_\chi/k_{\chi R})^2= F^{-4/3}$ \cite{Seto:2003kc}.
On the other hand, there are no effects on large scale 
with the mode $k<k_\chi$.

Thus the gravitational wave spectrum in the presence of 
late-time entropy production is completely characterized 
by two additional parameters, the dilution factor $F$ and 
the decay temperature of $\chi$, $T_\chi$. 
However, in many cases $T_\chi$ is expected to be very small, 
say, $\sim O(1)$~MeV-$O(1)$~GeV, 
and hence it does not affect the gravitational wave amplitude 
at the frequency relevant for the direct detection.
Thus hereafter we mainly focus on the effect of varying $F$.\par

Note that non-negligible $F$ affects only the overall amplitude 
of the gravitational wave background 
for the mode $k_{\chi R}<k<k_R$, and hence there is a degeneracy 
between $F$ and tensor-to-scalar ratio $r$ 
when concerning the direct detection around 0.1Hz.
However, the tensor-to-scalar ratio should be determined by 
cosmologically large scale CMB B-mode anisotropy. 
Thus if future CMB experiments measure $r$, 
there does not remain an ambiguity coming from 
the degeneracy between $r$ and $F$. In other words, 
if the result of direct detection deviates from the expected signal 
from the large scale measurement of $r$, 
there must be an entropy production process in the early universe. \par 

\begin{figure}[htbp]
 \begin{center}
   \includegraphics[width=0.6\linewidth]{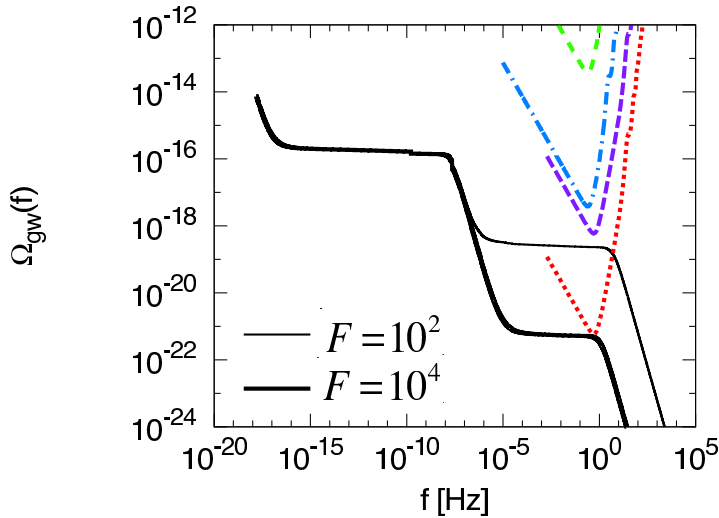} 
     \caption{Gravitational wave spectrum for the dilution factor $F=10^2$ and $10^4$.
     Here we have fixed $r=0.1$, $T_R=10^9$~GeV and $T_\chi=1$~GeV.}
      \label{fig:GWF}
 \end{center}
\end{figure}

All of these features are seen in Fig.~\ref{fig:GWF}, 
where the gravitational wave spectrum with $F=10^2$ and $10^4$ 
are shown. Here we have fixed $r=0.1$, $T_R=10^9$~GeV 
and $T_\chi=1$~GeV. This figure shows how the gravitational wave spectrum 
is affected by the late-time entropy production.

\begin{figure}[tbp]
 \begin{center}
   \includegraphics[width=0.6\linewidth]{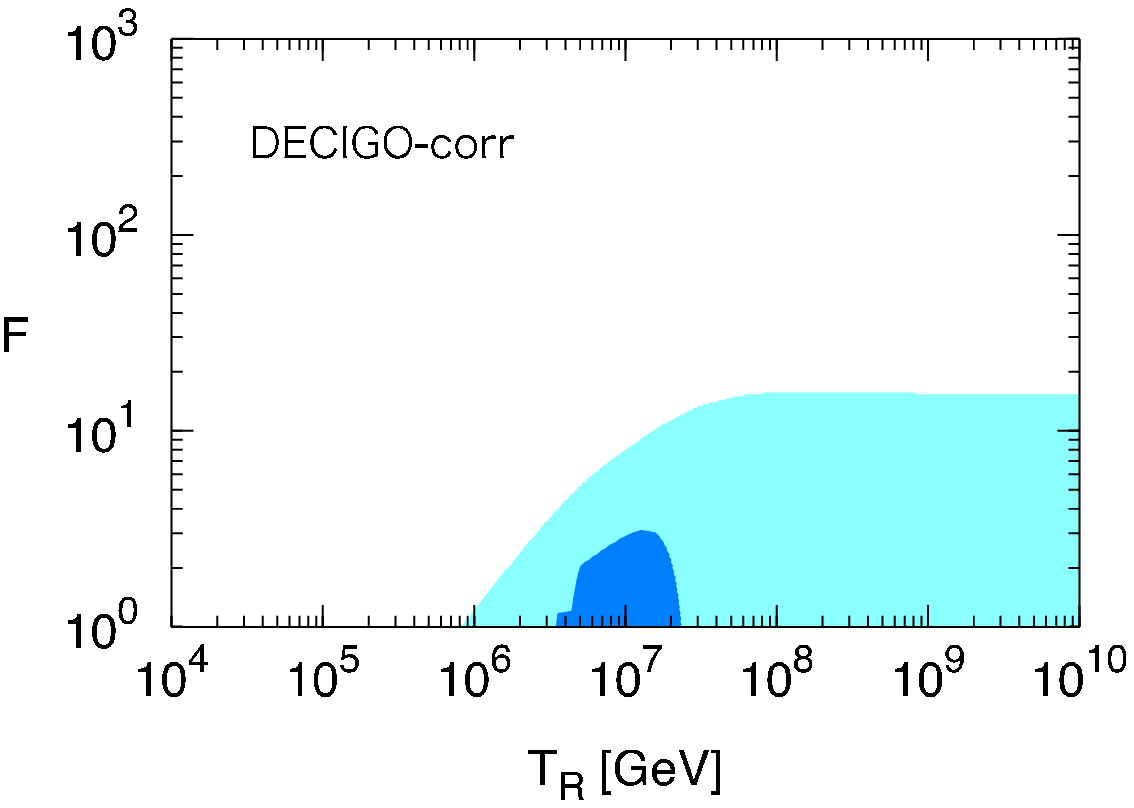} 
   \includegraphics[width=0.6\linewidth]{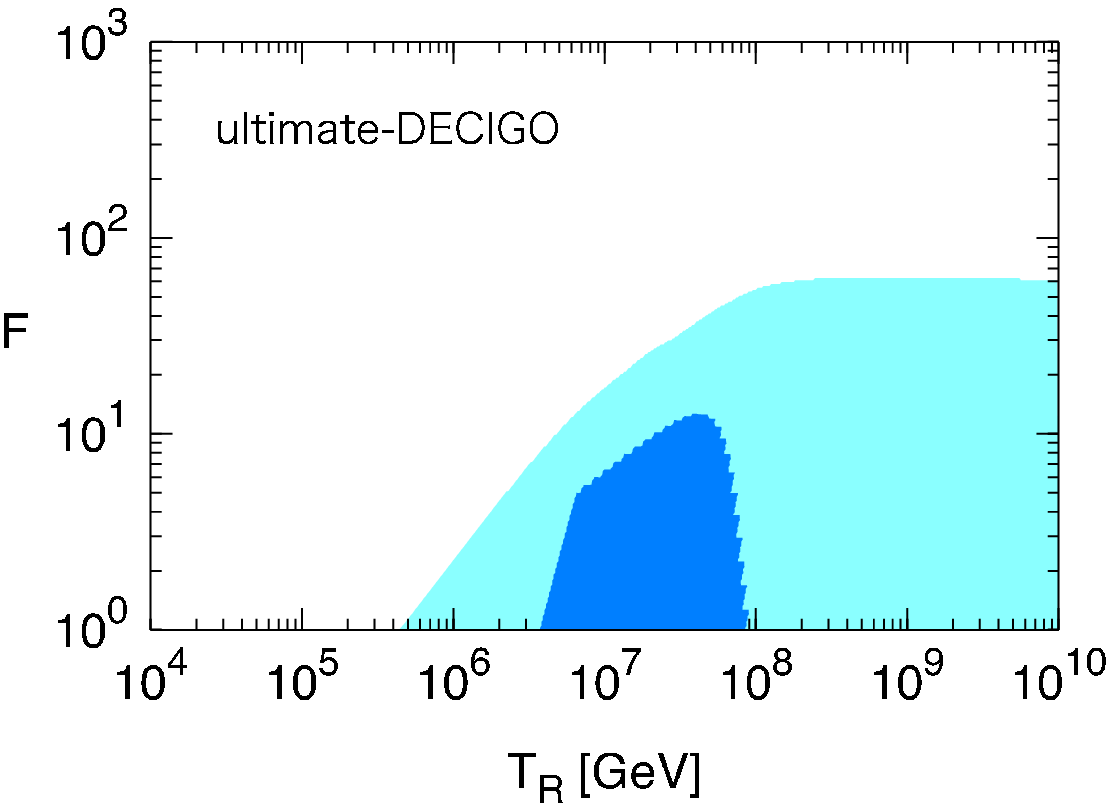} 
   \includegraphics[width=0.6\linewidth]{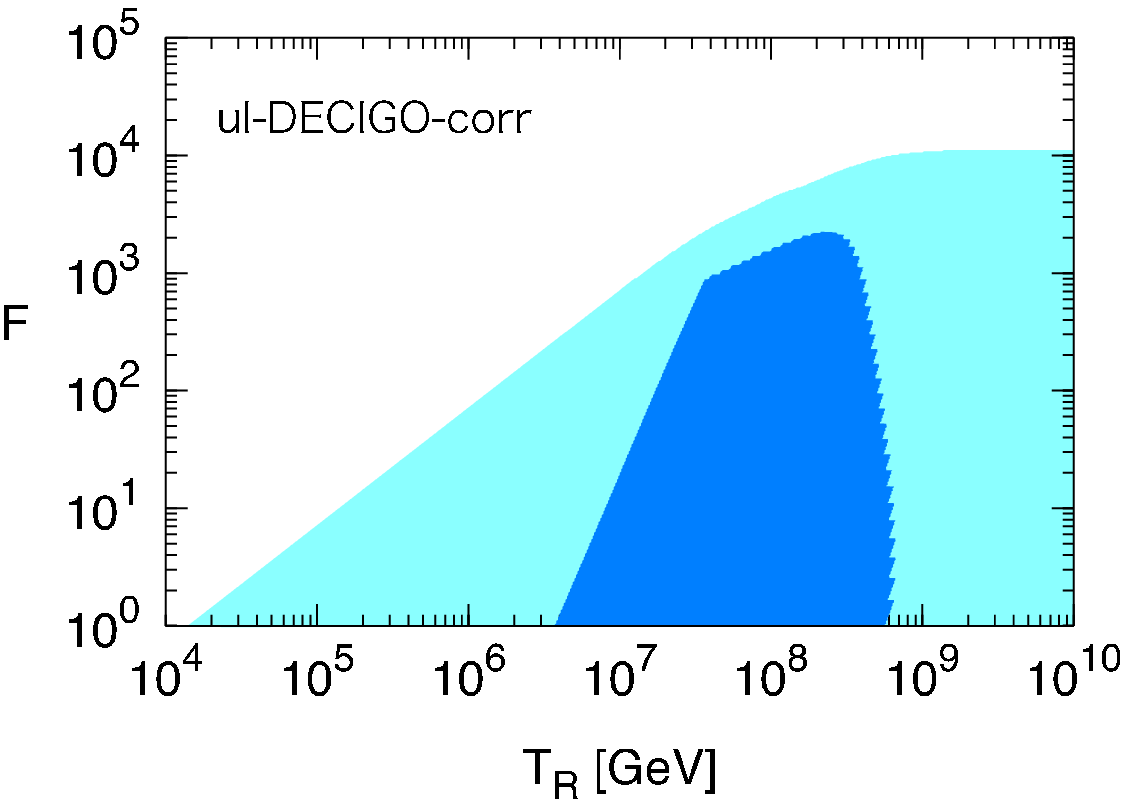} 
     \caption{In the outer light shaded region the gravitational wave background can be detected,
       and the inner blue shaded region shows the region where $T_R$ can be determined 
       with signal-to-noise ratio 5 by correlated analysis of DECIGO, ultimate-DECIGO (single)
       and ultimate-DECIGO (correlation) from upper to lower.
       Here $r=0.1$ is assumed.}
     \label{fig:FBBO}
 \end{center}
\end{figure}

If 0.1~Hz$\lesssim k_R(F) \lesssim 10$~Hz, 
both $F$ and $T_R$ can be determined from the shape of 
the gravitational wave spectrum. 
In Fig.~\ref{fig:GWF} we show future sensitivity 
to determine the dilution factor $F$ 
and the reheating temperature $T_R$ 
with fixed tensor-to-scalar ratio $r$, 
which can be measured by CMB polarization experiments. 
In Fig.~\ref{fig:FBBO} we show parameter 
region in $F-T_R$ plane where both parameters can be determined 
by correlated analysis of DECIGO, ultimate DECIGO (single),
and ultimate DECIGO (correlation) respectively by the dark shaded region.
The light shaded region shows the parameters where the gravitational wave background
is detected, but the value of $T_R$ remains undetermined.
Here we have fixed $r=0.1$ and $T_\chi =1$~GeV. 
(The precise value of $T_\chi$ does not matter 
as long as $T_\chi \lesssim 10^3$~GeV.) 
On the other hand if $k_R(F)\lesssim 0.1$~Hz, 
we can measure only the ratio $T_R/F$. However, this ratio contains 
sufficient information to determine the gravitino abundance including 
the dilution effect. Finally if $k_R(F)\gtrsim 10$~Hz, 
only a lower bound on $T_R$ is obtained as 
$T_R \gtrsim 2\times 10^9 F^{1/3}$~GeV. 
Even this case is useful for constraining the gravitino mass, 
as we will see. 

\section{Implications on particle physics}
\label{sec:implication}

We have shown that the reheating temperature of the universe 
$T_R$ may be determined 
from future space-based laser interferometer experiments. 
In this section we discuss some implications of determining $T_R$ 
on particle physics, in particular SUSY models\footnote{
	SUSY theories contain nearly twice particle species 
	compared with the standard model and the relativistic 
	effective degrees of freedom at high temperature is doubled, 
	$g_*(T\gtrsim 1~{\rm TeV}) =228.75$. 
	This leads to slight suppression on the gravitational wave spectrum, 
	but does not much affect the results of Sec.~\ref{sec:reheating}. 
}
and baryogenesis mechanisms.

\subsection{Supersymmetric models}

\subsubsection{Thermally produced gravitinos}

In SUSY, there exists the superpartner of the graviton, the gravitino. 
The gravitino mass ranges from $O(1)$~eV to $O(100)$~TeV 
depending on SUSY breaking models, 
while other SUSY particles have the mass of $O(1)$~TeV.
Gravitinos are produced in the early universe through scattering 
of particles in thermal bath, 
and the resultant abundance of the gravitino is proportional to $T_R$ 
\cite{Kawasaki:2004yh,Bolz:2000fu,Pradler:2006qh}, 
\begin{equation}
	Y_{3/2} \simeq 2\times 10^{-12} 
	\left (1+ \frac{m_{\tilde g}^2}{3m_{3/2}^2} \right )
	\left ( \frac{T_R}{10^{10}~{\rm GeV}} \right )
	\left ( \frac{1}{F} \right ), \label{TPgravitino}
\end{equation}
where $Y_{3/2}=n_{3/2}/s$ is the gravitino number-to-entropy ratio, 
$m_{3/2}$ denotes the gravitino mass 
and $m_{\tilde g} (\sim O(1)$~TeV) is the mass of the gluino, 
fermionic superpartner of the gluon.
If the gravitino is rather heavy $(m_{3/2}\gtrsim 1$~TeV) 
and unstable, it eventually decays with lifetime typically 
longer than 1 sec producing high-energy photons and hadrons. 
Those decay processes destroy or overproduce light elements 
such as {$^4$He}, D, {$^7$Li} and {$^6$Li}.
This constrains the reheating temperature as 
$T_R\lesssim 10^{6-9}$~GeV depending on the gravitino mass 
and its hadronic branching ratio \cite{Kawasaki:2004yh}.
On the other hand, if the gravitino is the lightest SUSY particle 
and stable due to the $R$-parity conservation, 
it contributes to total matter density 
of the universe \cite{Moroi:1993mb}. This leads to the constraint 
\begin{equation}
	T_R \lesssim 7\times 10^6~{\rm GeV}
	\left ( \frac{m_{3/2}}{1~{\rm GeV}} \right )
	\left ( \frac{m_{\tilde g}}{1~{\rm TeV}} \right )^{-2} F,
	\label{cons_TR}
\end{equation}
for $m_{3/2}\sim 10^{-4}$-$10$~GeV. \par

For example, let us consider the situation where 
$T_R$ is revealed to be larger than $\sim 10^7$~GeV 
and negligible dilution factor ($F\sim 1$) is confirmed 
with future space-based laser interferometer experiments.
For the unstable gravitino with $m_{3/2}\gtrsim 1$~TeV, 
as is often the case with gravity-mediated SUSY breaking models, 
its abundance is constrained as $Y_{3/2}\lesssim 10^{-16}$ 
\cite{Kawasaki:2004yh}, and hence $T_R$ should be less than 
around $\sim 10^6$~GeV. Thus SUSY breaking models which predict
the gravitino mass of $O(1)$~TeV will be excluded, 
even if the gravitino mass might not be determined 
by accelerator experiments. Also for the light gravitino scenario, 
the following constraint is obtained,  
\begin{equation}
	m_{3/2} \gtrsim 1~{\rm GeV} \left ( \frac{1}{F} \right )
	\left ( \frac{T_R}{10^7~{\rm GeV}} \right )
	\left ( \frac{m_{\tilde g}}{1~{\rm TeV}} \right )^{2}. 
	\label{cons_mgrav}
\end{equation}
Thus the gravitino mass with $m_{3/2}\lesssim 1$~GeV, 
which can be realized in some classes of gauge-mediated 
SUSY breaking models \cite{Giudice:1998bp}, is excluded 
since otherwise the gravitino abundance exceeds 
the present dark matter abundance. \par

On the other hand, if the gravitino is stable and
its mass is determined from accelerator experiments, 
$T_R$ is the only parameter which determines 
the present gravitino abundance. 
If the observationally inferred value of $T_R$ saturates 
the upper bound (\ref{cons_TR}), the gravitino abundance 
is appropriate for the dark matter of the universe.\footnote{
	 Gravitinos produced by the decay of the next-to-lightest SUSY particle may also be
	 the dark matter \cite{Roszkowski:2004jd,Steffen:2006hw},
	 but BBN constraints almost exclude such a possibility.
}
Note that if the dark matter consists of the gravitino, 
it is impossible to detect it directly 
because of its too weak interaction strength. 
Thus $T_R$ has an important information to probe 
whether the gravitino truly takes a roll 
of the dark matter or not. \par

\subsubsection{Non-thermally produced gravitinos}

Recently it is pointed out that non-thermal production process of gravitinos from the inflaton decay 
gives a significant amount of gravitino abundance \cite{Kawasaki:2006gs},
\begin{equation}
	Y_{3/2}^{(\rm NT)} \simeq 9\times 10^{-11} 
	\left ( \frac{m_\phi}{10^{13}~{\rm GeV}} \right )^{2}
	\left ( \frac{\langle \phi \rangle}{10^{15}~{\rm GeV}} \right )^{2}
	\left ( \frac{10^{6}~{\rm GeV}}{T_R} \right ),  \label{NTgravitino}
\end{equation}
where $m_\phi$ and $\langle \phi \rangle$ denote the mass and VEV of the inflaton.
Since non-thermal contribution is proportional to $T_R^{-1}$, this gives lower bound on $T_R$
for fixed $m_{3/2}$.
Furthermore the inflaton always decay into MSSM sector appearing in the superpotential 
in supergravity, which provides lower limit on the reheating temperature as \cite{Endo:2006qk}
\begin{equation}
	T_R \gtrsim 10~{\rm TeV} |y_t|
	\left ( \frac{228.75}{g_*(T_R)} \right )^{1/4}
	\left ( \frac{m_\phi}{10^{13}~{\rm GeV}} \right )^{3/2}
	\left ( \frac{\langle \phi \rangle}{10^{15}~{\rm GeV}} \right ), \label{lowbound}
\end{equation}
where $y_t$ is the top Yukawa coupling.

\begin{figure}[tbp]
 \begin{center}
   \includegraphics[width=0.7\linewidth]{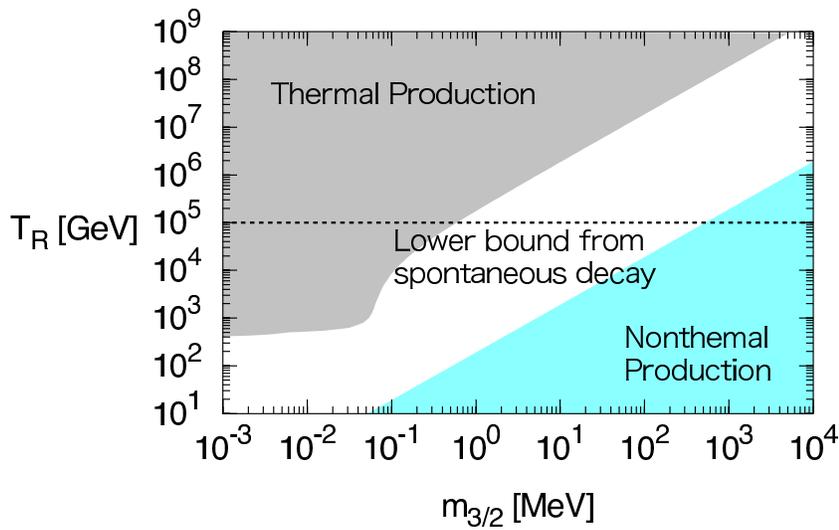} 
     \caption{Constraints on the reheating temperature against the gravitino mass. 
     The upper left region is excluded from thermal production, and lower right region is
     excluded from non-thermal production for $\langle \phi \rangle=10^{15}$~GeV 
     and $m_{\phi}=10^{13}$~GeV.
     Also spontaneous decay of the inflaton restricts $T_R \gtrsim 10^5$~GeV 
     as denoted by the dotted line. }
     \label{fig:m3/2-TR}
 \end{center}
\end{figure}

In Fig.~\ref{fig:m3/2-TR} allowed parameter region in $m_{3/2}$-$T_R$ is shown
for the case of stable gravitino
for $m_\phi=10^{13}~{\rm GeV}$ and $\langle \phi \rangle = 10^{15}~{\rm GeV}$.
The upper left and lower right regions are excluded from thermal (\ref{TPgravitino})
and nonthermal production (\ref{NTgravitino}), respectively.
Also spontaneous decay of the inflaton gives lower bound on $T_R$ (\ref{lowbound})
as denoted by the dotted line.
It is seen that $10^5~{\rm GeV} \lesssim T_R \lesssim 10^9~{\rm GeV}$ is favored,
and this range is interesting from the viewpoint of direct detection of gravitational waves.

\subsection{Baryogenesis mechanism}

Another important issue which is deeply related to 
the reheating temperature is baryogenesis. 
As is well known, almost all the ordinary matter 
in the universe consists of baryon, not anti-baryon. 
It is a long-standing mystery how the sizable amount of baryon asymmetry 
is generated after inflation. We must rely on some new physics 
which involves extra baryon number violation, 
CP violation and non-equilibrium process in order to 
create matter-anti-matter asymmetry. 
One of the most popular mechanisms to create a correct amount 
of baryon asymmetry is thermal leptogenesis scenario 
using right-handed neutrino \cite{Fukugita:1986hr}. 
In order for this mechanism to work well, 
$T_R \gtrsim 10^9$~GeV is required \cite{Buchmuller:2005eh}. \par

Right-handed (s)neutrinos can also be produced non-thermally 
via right-handed sneutrino condensation 
\cite{Murayama:1992ua,Murayama:1993em} 
or inflaton decay \cite{Lazarides:1991wu,Asaka:1999yd}, 
and $T_R\gtrsim 10^6$~GeV is required in order to create 
observed amount of baryon asymmetry in these scenarios. 
Thus non-thermal leptogenesis scenarios can also be 
favored or disfavored from 
future space-based gravitational wave detectors. \par

Besides those leptogenesis scenarios, 
there are many other baryogenesis mechanisms 
which we do not list here \cite{Dine:2003ax}, 
and many of them predicts baryon asymmetry proportional to $T_R$. 
Thus determining $T_R$ has important implications 
on baryogenesis mechanisms.

\section{Conclusions and discussion}
\label{sec:conclusion}

In this paper we have shown that direct detection of stochastic 
gravitational wave background 
of inflationary origin carries rich information on the early universe.
In particular, the reheating temperature of the universe after 
inflation $T_R$ can be determined 
with future space-based laser interferometer experiments, DECIGO and BBO. 
In case that non-zero tensor-to-scalar ratio is confirmed 
by CMB polarization observations, 
 DECIGO or BBO will detect signals of gravitational wave background 
if $T_R \gtrsim 10^5$~GeV. 
Moreover, they can not only detect gravitational waves, 
but determine $T_R$ if $T_R\lesssim 10^9$~GeV, 
through the $T_R$ dependence of gravitational wave spectrum. 
Thus combined analysis of direct detection of gravitational 
wave background and CMB anisotropy measurements provide consistency 
check of inflation models. 
It also restricts realistic model among many and many inflation models, 
since both the tensor-to-scalar ratio and the reheating 
temperature crucially depend on the properties of the inflaton 
- its mass, potential, interaction strength, etc. \par

We have also discussed implications of determination 
of the reheating temperature on particle physics.
The detection of gravitational wave background clearly 
goes beyond just a probe of thermal history before BBN.
In the environment of the very early universe 
with extremely high temperature, currently undiscovered particles 
predicted by some physics beyond the standard model 
may be efficiently produced. 
Thus probing this epoch is directly  connected to particle physics. 
For example, some SUSY breaking models will 
be excluded or severely constrained from the so-called gravitino problem, 
if future observations determine $T_R$. 
Also some baryogenesis scenarios, including thermal 
leptogenesis scenario, may be excluded or disfavored.
In SUSY axion models, constraints on $T_R$ may become much more stringent 
\cite{Covi:2001nw,Kawasaki:2007mk}.\par 

As a final remark, gravitational waves may also be generated 
from some other cosmological processes, such as preheating 
after inflation 
\cite{Khlebnikov:1997di,Easther:2006gt,GarciaBellido:2007dg}
and first order phase transition followed by subsequent 
bubble collisions \cite{Kosowsky:1991ua,Easther:2008sx}. 
Although the amplitude and typical frequency from these contributions 
are highly model-dependent, they might give complementary 
information on inflation models or cosmological evolution scenario 
after inflation, if detected independently of primordial gravitational 
waves analyzed in this paper. 
Moreover, recently it is pointed out that decay of domain walls 
associated with gaugino condensation 
may produce a significant amount of gravitational waves and 
can be used as a probe of the gravitino mass \cite{Takahashi:2008mu}.

\ack{
The authors are grateful to Naoki Seto, Fuminobu
Takahashi, Atsushi Taruya, and Asantha Cooray for useful
communications.  This work was partially supported by JSPS through
research fellowships (KN, YS) and Grant-in-Aid for Scientific Research
No.~ 19340054(JY).
}


\section*{References}



\end{document}